\documentclass{article} 
\usepackage{iclr2023_conference,times}

\usepackage{amsmath,amsfonts,bm}

\def\eqref#1{equation~\ref{#1}}

\def\1{\bm{1}}

\DeclareMathAlphabet{\mathsfit}{\encodingdefault}{\sfdefault}{m}{sl}
\SetMathAlphabet{\mathsfit}{bold}{\encodingdefault}{\sfdefault}{bx}{n}

\usepackage{subcaption}
\usepackage{microtype}
\usepackage{booktabs} 
\usepackage{multirow}
\usepackage{tabularx}
\usepackage{xcolor}
\usepackage{url}
\usepackage{caption}
\usepackage{tabularx}
\usepackage{multicol}
\usepackage{floatrow}
\usepackage{microtype}
\usepackage{amsbsy}
\usepackage{amsmath}
\usepackage{xspace}
\usepackage{inconsolata}
\usepackage{amssymb}
\usepackage{adjustbox}
\usepackage{textcomp}
\usepackage{color, colortbl}
\usepackage{arydshln}
\usepackage{pifont}
\usepackage{url}
\usepackage[colorlinks=true,allcolors=blue]{hyperref}

\usepackage[LGR,T1]{fontenc}
\usepackage[greek,english]{babel}

\newcommand{\model}{{\textsc{Pheme}}\xspace}
\newcommand{\mqtts}{{\textsc{mqtts}}\xspace}

\newcommand{\rparagraph}[1]{\vspace{1.2mm}\noindent\textbf{#1.}}

\newcommand{\sparagraph}[1]{\vspace{0.0mm}\noindent\textbf{#1.}}

\usepackage{todonotes}
\makeatletter
\newcommand*\iftodonotes{\if@todonotes@disabled\expandafter\@secondoftwo\else\expandafter\@firstoftwo\fi}
\makeatother

\definecolor{Gray}{gray}{0.92}
\newcolumntype{Y}{>{\centering\arraybackslash}X}

\title{\model: Efficient and Conversational Speech Generation}

\author{Paweł Budzianowski, Taras Sereda, Tomasz Cichy, Ivan Vuli\'{c}\\
PolyAI Limited\\
London, United Kingdom\\
\texttt{\{pawel,ivan\}@poly-ai.com}}

\iclrfinalcopy 
\begin{document}

\maketitle

\begin{abstract}
In recent years, speech generation has seen remarkable progress, now achieving one-shot generation capability that is often virtually indistinguishable from real human voice. Integrating such advancements in speech generation with large language models might revolutionize a wide range of applications. However, certain applications, such as assistive conversational systems, require natural and conversational speech generation tools that also operate efficiently in real time. Current state-of-the-art models like VALL-E and SoundStorm, powered by hierarchical neural audio codecs, require large neural components and extensive training data to work well. In contrast, MQTTS aims to build more compact conversational TTS models while capitalizing on smaller-scale real-life conversational speech data. However, its autoregressive nature yields high inference latency and thus limits its real-time usage. In order to mitigate the current limitations of the state-of-the-art TTS models while capitalizing on their strengths, in this work we introduce the \model model series that \textbf{1)} offers compact yet high-performing models, \textbf{2)} allows for parallel speech generation of \textbf{3)} natural conversational speech, and \textbf{4)} it can be trained efficiently on smaller-scale conversational data, cutting data demands by more than 10x but still matching the quality of the autoregressive TTS models. We also show that through simple teacher-student distillation we can meet significant improvements in voice quality for single-speaker setups on top of pretrained \model checkpoints, relying solely on synthetic speech generated by much larger teacher models. Audio samples and pretrained models are available online.
\end{abstract}

\section{Introduction}
\label{sec:intro}
A crucial requirement for wider adoption of many modern AI applications, and especially conversational systems such as voice assistants, is the ability to synthesise natural, human-level or `lifelike' speech~\citep{Latif:2023survey,m2ctts}. Owing to the use of deep learning architectures, neural text-to-speech (TTS) synthesis has seen some tremendous improvements \citep{vits,valle,speartts,instructtts,styletts}, and the pace of development has been further accelerated recently. In particular, the progress has been unlocked by casting speech processing from a continuous to a discrete domain via \textit{neural audio codecs} \citep{encodec,soundstream,beats,mqtts,tok-vec}: this enables the use of standard and powerful Transformer-based sequence-to-sequence modeling paradigms \citep{transformers,t5}, established in the work on large language models, for audio generation. Such Transformer-based architectures have been successfully applied to text-to-speech synthesis \citep{audiolm,valle,speartts,soundstorm,speechtokenizer}, as well as to music generation \citep{musiclm} and generation of other audio signals \citep{audiogen,audiogpt}.

\textit{Conversational TTS} is \textit{`condition sine qua non}' of the lifelike user experience that would increase the overall satisfaction of users interacting with assistive conversational systems. However, the majority of current work on Transformer-based TTS still trains and evaluates models in controlled `upper bound' environments that rely on cleanly recorded reading (e.g., audiobooks) or acted speech. Such environments effectively ignore the undeniable fact that human speech acts across many scenarios are conversational in nature, and that the TTS systems must be highly adaptable to applications across many different domains and should also be useful `in the wild' (e.g., over the phone). Moreover, conversational TTS synthesis poses additional challenges as conversational speech conveys additional paralinguistic information (e.g., emotions, pitch), and is much richer in terms of prosodic variations and expressiveness \citep{denoispeech,adaspeech4,m2ctts}.

Another key practical imperative is the \textit{efficiency of the TTS systems}~\citep{efficienttts}, where efficiency covers the following pivotal aspects. First, parameter efficiency requires the TTS models to be compact and thus shareable and runnable on consumer hardware. Second, data efficiency requires the models to learn high-quality and competitive models for particular scenarios and voices even with limited amounts of in-domain or in-voice data (or even in `zero-shot' or `one-shot' prompting setups). In addition, smaller-scale pretraining data would enable quicker development cycles. Finally, inference efficiency (with low latency) makes the models production-friendly and usable in `real-life' applications (e.g., as a component of conversational assistants speaking over the phone).

In this work, we first take stock of the current progress in Transformer-based neural TTS models (see later in \S\ref{sec:rw}): we start by detecting the limitations of the current models and then, inspired by the previous work, adopt and adapt their desirable properties and features during the construction of our efficient and conversational Transformer-based TTS system: PolyAI's \model.\footnote{In Greek mythology, \model is the goddess and personified spirit of, among others, fame and rumor and is described as `she who initiates and furthers communication'. Moreover, the word \model is related to the Greek word $\phi$\textgreek{άναι} meaning \textit{`to speak'}.} The main goal of \model is to maintain high-quality TTS both in multi-speaker and single-speakers scenarios, while simultaneously providing the following features: \textbf{1)} synthesis of rich-in-prosody and naturally sounding speech rather than acted and artificial speech; \textbf{2)} compact, shareable and easily fine-tunable models which also work with unseen voices through one-shot prompting; \textbf{3)} reduced pretraining time and data requirements; \textbf{4)} high inference efficiency and low latency. Further, as another contribution, we propose and empirically validate that further gains in single-speaker specialization scenarios could be achieved via leveraging synthetic data created by much larger models or 3rd-party software. 

Our experiments assess all the key properties of \model such as speech intelligibility, naturalness and diversity, as well as inference efficiency. The main comparison with the previous state-of-the-art conversational TTS model, MQTTS~\citep{mqtts},\footnote{MQTTS also showed improved performance on conversational TTS in comparison to some non-autoregressive TTS models such as VITS~\citep{vits}, and is in spirit closest to our work.} reveal that \model offers improved intelligibility and naturalness at the same or even reduced model size, while dramatically increasing inference speed via incorporating non-autoregressive parallel decoding, and this way providing true `real-time' synthesis capabilities. Owing to this plus its design simplicity, \model might serve as a solid framework for further developments in conversational and efficient TTS, and in hope to guide such developments we release our code and pretrained models online at \url{https://polyai-ldn.github.io/pheme/}.

\section{Related Work and Background}
\label{sec:rw}
This work builds upon a recent stream of research on the so-called Transformer-based \textit{large audio models}, where the key idea is to apply large language models (LLMs) in the field of audio processing and generation. For a broader overview, we refer the reader to a recent comprehensive survey \citep{Latif:2023survey}, while here we briefly survey prior work related to conversational and efficient TTS, and key modeling choices for \model.

\sparagraph{Discrete Speech Representations: Semantic and Acoustic Tokens}
Neural audio codecs \citep[\textit{among others}]{encodec,soundstream,beats,speechtokenizer} provide the capability of reconstructing high-quality audio at very low bitrates, which in turn enables speech modeling in the discrete domain (e.g., through standard sequence-to-sequence LLM-style paradigms). Prior work typically divides the discrete speech representations into two core groups: \textbf{1)} \textit{semantic tokens} and \textbf{2)} \textit{acoustic tokens}. This duality aims to disentangle the content related to pure semantics (which should be fully captured by semantic tokens) from the paralinguistic content (e.g., speaker identity, prosody, timbre) which should be captured by acoustic tokens. Ideally, the full disentaglement would enable fully separate development of the two core system components: \textbf{1)} \textit{text-to-semantics} (T2S) aiming to learn to generate the layer of semantic tokens and \textbf{2)} \textit{acoustics-to-speech} (A2S) which learns to generate speech output from the generated acoustic tokens which were conditioned on the (previously generated) semantic tokens. This particular design is at the heart of the most recent models such as SpearTTS~\citep{speartts}, MQTTS~\citep{mqtts}, SoundStorm~\citep{soundstorm}, and USLM~\citep{speechtokenizer}, and we thus also adopt this disentanglement into T2S and A2S modeling in this work.

\sparagraph{Residual Vector Quantization for TTS}
While there are some technical differences between different TTS architectures related to how semantic and acoustic tokens are constructed, on a high level the majority of recent work relies on the concept of \textit{residual vector quantisation (RVQ)}.\footnote{For a nice and didactic overview, we refer the reader to the following write-up: \url{https://drscotthawley.github.io/blog/posts/2023-06-12-RVQ.html}.} Here, each compressed audio frame gets quantized by multiple hierarchical layers of quantizers, where each subsequent layer works with the residual of the previous layer. For instance, SoundStorm borrows a hierarchical sequence-to-sequence approach of AudioLM~\citep{audiolm} where semantic tokens are used as an intermediate conditioning signal to generate acoustic tokens from the SoundStream codec~\citep{soundstream}. Other models that use RVQ include VALL-E~\citep{valle}, RVQGAN~\citep{rvqgan}, NaturalSpeech 2~\citep{naturalspeech2}, MQTTS~\cite{mqtts}, among others.

However, the work of \citet{speechtokenizer} has empirically validated that the disentanglement desiderata between semantic and acoustic information in respective semantic and acoustic tokens are often not fully met (i.e., there are undesirable residuals of semantic information in higher quantization layers). They have thus proposed a unified SpeechTokenizer RVQ method which creates $Q$ quantization layers, where the first layer is treated as the layer of semantic tokens, while all the other $Q-1$ layers are treated as hierarchical layers of acoustic tokens. SpeechTokenizer shows a higher rate of disentanglement than some other standard choices which is why we use this as our go-to model to derive RVQ-based semantic and acoustic tokens. We discuss this in more detail in \S\ref{sec:methodology}.

\sparagraph{Conversational TTS}
As mentioned before, the majority of prior TTS work has focused on reading or acted speech environments and datasets such as LibriSpeech~\citep{librispeech} or VCTK.\footnote{\url{https://datashare.ed.ac.uk/handle/10283/3443}} VALL-E trains on a much larger LibriLight dataset~\citep{librilight} covering 60k hours of speech coupled with transcriptions generated by an automatic speech recognition (ASR) system. Concerning truly conversational TTS, SoundStorm~\citep{soundstorm} trains a dialogue synthesis model on a \textit{cleaned and publicly \underline{unavailable}} corpus of approximately 100,000 hours of dialogue. 

In this work, following the relevant prior work on conversational TTS, MQTTS, we adopt and use the GigaSpeech corpus~\cite{gigaspeech}, which in its largest, `XL' variant covers 10k hours of noisy speech. Similar to LibriLight, GigaSpeech is also an `ASR-ed, silver transcriptions' corpus which spans transcribed audio from audiobooks, podcasts, and YouTube with a 16 kHz sampling rate. This effectively creates a noisy conversational dataset, where noise might come from imperfect speech transcriptions as well as from the actual speech sources with occasionally ill-defined speech. Our primary goal is to demonstrate that it is possible to pretrain good-quality conversational TTS models even with such noisy data (or at least after filtering noise from the data, see later in \S\ref{sec:experiments}) which is more than 10x smaller than what SoundStorm has been trained on, and more than 6x smaller than the audiobooks-only pretraining dataset used by VALL-E.





\rparagraph{Efficient TTS}
Next, in order to satisfy the efficiency criteria, we rely on the following findings from prior work. First, the real-time usage of the current state-of-the-art conversational TTS model, MQTTS, is limited by its autoregressive nature. In order to boost efficiency, we instead adopt the recently proposed masking (non-autoregressive) parallel decoding scheme of MaskGIT~\citep{maskgit}, further adapted to speech token sequences produced by RVQ in the work of \citet{soundstorm}. In a nutshell, MaskGIT starts from masked tokens and then, across multiple iterations, predicts a portion of the tokens based on their confidence scores, where the portion typically increases during the iterative procedure progression. Technical details of MaskGIT-style decoding for \model are discussed in \S\ref{sec:experiments}, and in \S\ref{sec:results} we show that the parallel decoding scheme does not compromise generation quality even when applied with very compact models such as \model while offering huge boosts in inference speed.

Regarding parameter efficiency, our T2S and A2S components remain modest in size, again following the MQTTS design principle. For instance, we limit the T2S component (see later in \S\ref{sec:methodology}) to a standard T5-style~\citep{t5} sequence-to-sequence model with a maximum of 50M or 100M parameters.  Further details on \model model size are covered in \S\ref{sec:methodology} and \S\ref{sec:experiments}. In comparison, SpearTTS uses by an order of magnitude larger T5-style T2S model (500M+ parameters) while SoundStorm relies on a ByT5-Large model~\citep{byt5} for T2S: there, solely the T2S component already stretches to 1.23B parameters, which makes the model very cumbersome in production. Similarly, regarding data efficiency, for their T2S component SpearTTS relies on very expensive unsupervised pretraining plus additional 60k hours of LibriLight data with an additional back-translation step; SoundStorm learns the T2S component using the curated corpus of 100k hours of spoken dialogue. Here, we aim to learn the T2S component again relying solely on supervised training (without any time-consuming pretraining and back-translation) on much smaller datasets, as discussed before.

Finally, as another efficiency aspect, following the rich body of work on distillation for speech models \citep{huang:2023ensemble,dphubert}, we investigate if efficient and high-quality specialized single-speaker models can be obtained (i.e., fine-tuned) from pretrained \model checkpoints, relying solely on modest amounts (e.g., 10h or less) of synthetic speech from larger TTS models. 



\section{Methodology}
\label{sec:methodology}
In what follows, we now delve deeper into the main components of \model, covering \textbf{1)} speech tokenization, \textbf{2)} learning of the T2S component and \textbf{3)} parallel non-autoregressive decoding for the A2S component, showcasing its simple and efficient design.

\sparagraph{Speech Tokenization}
As the source of semantic and acoustic tokens, we rely on the recent SpeechTokenizer model~\citep{speechtokenizer}.\footnote{\url{https://github.com/ZhangXInFD/SpeechTokenizer}} It operates on a single-channel audio signal represented as a sequence $x \in \mathbb{R}^T$, where $T$ denotes the length of the input sequence, $T=d\cdot r_s$, where $d$ denotes duration and $r_s$ is the sampling rate (e.g., 16 kHz in GigaSpeech). For simplicity, we use the available off-the-shelf variant of the SpeechTokenizer which was pretrained on LibriSpeech, without any additional adaptive fine-tuning on GigaSpeech. The full implementation and training details of SpeechTokenizer are available in the original work.

SpeechTokenizer, unlike other RVQ-based speech tokenization methods, unifies semantic and acoustic tokens, aiming to better disentangle different aspects of speech information hierarchically across the corresponding RVQ layers.\footnote{The importance of this disentanglement was also evidenced in our preliminary experiments. Besides SpeechTokenizer, we (i) investigated the standard approach of using $C$ k-means centroids of the embeddings extracted from layers 7 or 9 of the HuBERT model \citep{hubert} as the layer of semantic tokens, coupled with RVQ-based acoustic tokens obtained by DAC \citep{rvqgan,vampnet} and EnCodec \citep{encodec}. In order to match the sampling rates between the two heterogeneous sources of semantic and acoustic tokens, we either duplicated neighboring semantic tokens or stretched audio we supplied as input to the HuBERT model. Regardless of the chosen model combination or alignment strategy, the preliminary experiments revealed much weaker, subpar performance when compared to the finally selected variant based on SpeechTokenizer.} The final output comprises $Q=8$ hierarchical RVQ layers $q_1,\ldots,q_8$ with the codebook size of $C=1,024$ in each RVQ layer, that is, it is a matrix $\{1,\ldots,C\}^{Q \times T}$. The row/layer $q_1$ is taken as the layer of \textit{semantic tokens} (for learning the T2S component and conditioning the A2S component), and the remaining RVQ layers $q_2$-$q_8$ are treated as hierarchical \textit{acoustic tokens}. 

\sparagraph{T2S: Training and Inference}
The first critical functionality for TTS is learning the mapping (or rather a sequence-to-sequence `translation') from the input raw text to the representation in the layer of semantic tokens: the output of the T2S component in practice is a $T$-dimensional vector of integers sampled (with repetition) from the $|C|$-item set $\{1,\ldots,C\}$. We treat T2S as the standard sequence-to-sequence problem and train a T5-style encoder-decoder architecture from scratch. 

The raw text $t_r$ is first preprocessed into the sequence of IPA phones relying on the standard IPA phonemizer: $t_{ipa} = \text{IPA}(t_r)$, where $\text{IPA}()$ denotes the phonemization function.\footnote{We use \url{https://github.com/bootphon/phonemizer}.} The T5 model architecture is initialized from scratch with random weights and a specialised vocabulary of 1,119 items in total. The vocabulary comprises the 1,024 different codes (i.e., integers from 1 to 1024) from the codebook of SpeechTokenizer, plus all the phones from the IPA phonemizer, plus the special tokens for begin-of-sequence and end-of-sequence.

In the basic form, the T5 model learns the following T2S function: $s_t = \text{T2S}(t_{ipa})$, where $s_t$ is the resulting vector of semantic tokens for the target text. The ground truth semantic tokens are obtained by applying SpeechTokenizer directly on training data allowing for duplicated tokens. This is the format we use during training. 

At inference, we can additionally prompt the \model model by providing the speech prompt $x_{p}$ along with its text transcription $t_p$: we run the prompt through SpeechTokenizer to obtain its semantic tokens $s_p$ (i.e., $q_1$ output from the tokenizer) and 7 levels of acoustic tokens (denoted as $a_{p,1},\ldots,a_{p,7}$, which correspond to RVQ output layers $q_2$-$q_8$ from the tokenizer). At inference, we then provide the concatenation of [$t_{ipa,p}$,$t_{ipa,t}$] as the full input to the T2S component, where $t_{ipa,p}$ is the IPA-based representation of $t_p$ and $t_{ipa,t}$ is the IPA-based representation of the target text for which we want to generate speech. When generating semantic tokens, we then also directly provide $s_p$ as the incomplete, first part of the full output, and let the model generate the rest of the output sequence, that is, $s_t$ - the sequence of semantic tokens for the target text.

We acknowledge that this approach to T2S is only one plausible option among other potential variants, such as the ones that preprocess input text data into CMU-based sequences of phonemes~\citep{speartts} or apply byte-level models directly on raw text $t_r$~\citep{soundstorm}; we leave other variants for future exploration.

\sparagraph{A2S: Training and Inference}
\label{ss:a2s}
We adopt the non-autoregressive procedure of SoundStorm, which is in turn an extension of masking and confidence-based parallel decoding scheme of MaskGIT~\citep{maskgit}. The RVQ-based SpeechTokenizer by default provides the coarse-to-fine order of the RVQ hierarchy, which is a crucial requirement. As discussed by \citet{soundstorm}, this means that \textbf{1)} the conditional dependencies between different RVQ levels are taken into account during training and inference, and \textbf{2)} sampling of finer levels, given all the tokens from coarser levels, can be conducted in parallel without loss of quality: they are responsible to capturing very local and fine-grained acoustic details, typically independent of each other. We further extend the SoundStorm method via the use of SpeechTokenizer-obtained acoustic tokens, and via the introduction of speaker embeddings in the pipeline. A brief description is provided in what follows, and we refer the reader to prior work~\citep{maskgit,soundstorm} for further technical details.

\noindent \textit{Masking.}
First, a time step $t \in \{1,\ldots,T-1\}$ is sampled uniformly at random, acting as the \textit{prompt delimiter}. We do not mask any tokens before the prompt delimiter, effectively preparing the model for the `voice prompting' setup at inference. Furthermore, we never mask the so-called \textit{conditioning tokens} comprising the full level of semantic tokens ($q_1$ from SpeechTokenizer at training or obtained via the T2S component at inference). Furthermore, we propose to use speaker embeddings at all $T$ time steps as additional conditioning tokens: they are obtained via a state-of-the-art speaker diarization model \texttt{pyannote} \citep{pyannote1}.\footnote{\url{https://github.com/pyannote/pyannote-audio}} The idea is that the direct use of speaker embeddings would yield higher generation fidelity than letting the model use the speaker information only implicitly.

Next, the current RVQ level $q_c$ is sampled uniformly at random from the set $\{q_2,\ldots,q_8\}$. Following that, the binary $T$-dimensional mask $M$ is sampled according to a cosine schedule for the current level $q_c$. More formally, the masking probability is $p=cos(u)$ where $u$ is uniformly sampled from the interval $[0,\pi/2]$, and $M_i$ is then sampled from the Bernoulli distribution with the probability $p$. Following that, we mask the selected non-prompt tokens at $q_c$ for which it holds: (i) they appear at a time step $t_m$ where $t_m > t$, (ii) $M_{t_m} = 1$. Finally, we mask all non-prompt tokens at finer RVQ levels, that is, it holds that (i) they appear at a time step $t_m>t$ and (ii) their associated level $q > q_c$.

\noindent \textit{Training.}
After obtaining the masked token sequence, the A2S model is trained via the standard cross-entropy loss, where the loss is calculated on the masked tokens of the level $q_c$ (and not on tokens from levels $q>q_c$). The A2S model is a standard Conformer network with bidirectional self-attention~\citep{conformer} and rotary positional embeddings~\citep{rope}.

\noindent \textit{Decoding.}
Given the conditioning tokens and (non-mandatory) prompt tokens, the iterative parallel decoding scheme is exactly the same as with SoundStorm: it proceeds RVQ level-wise, moving to the next RVQ level $q+1$ only after all the tokens from all the previous levels $1,\ldots,q$ have been sampled. Within each RVQ level, we rely on the confidence-based sampling regime of \citep{maskgit}. In summary, this level-wise iterative and parallel procedure substantially decreases the number of required forward passes in comparison to fully autoregressive models such as MQTTS; see again \citep{soundstorm} for other technical details.

This design with speaker embeddings then allows both for \textbf{1)} \textit{one-shot} speech generation: we provide both the short prompt with or without the speaker embedding to the model at inference; \textbf{2)} \textit{zero-shot} generation: we do not supply the prompt but provide the speaker embedding to the A2S component.

\section{Experimental Setup}
\label{sec:experiments}
\sparagraph{Model Configurations}
We train \model in two different sizes, labeled \textsc{small} and \textsc{large}. First, the 100M parameter model (with $\sim$45M parameters allocated to the T2S component, and $\sim$55M allocated to the A2S component) serves to run fair comparisons against the relevant prior work, MQTTS, where both models are trained solely on filtered conversational data from GigaSpeech (see \S\ref{ss:setups}). Second, the 300M model ($\sim$100M + $\sim$200M) aims to measure how well \model can scale with more parameters and data provided (including non-conversational data), and how its increase in size impacts its inference efficiency. The detailed model parameters are provided in Appendix~\ref{app:model_parameters}.

\subsection{Training and Inference Setups}
\label{ss:setups}
\sparagraph{Training Data}
For training the smaller, 100M \model variant, we use a filtered and preprocessed version of GigaSpeech~\citep{gigaspeech}, where we remove a large number of ill-defined and extremely noisy speech, following prior work~\citep{mqtts}. In particular, we keep only utterances from podcasts and YouTube portions of GigaSpeech XL, with estimated signal-to-noise-ratio (SNR) $>$20 dB and with duration between 5 and 15 seconds.  After the filtering step, we resample all utterances to 16 kHz and normalize loudness to -20 dB. For the SNR estimation we use \texttt{WADA-SNR}\footnote{\url{https://gist.github.com/johnmeade/d8d2c67b87cda95cd253f55c21387e75}} and for loudness normalization we use \texttt{pyloudnorm}\footnote{\url{https://github.com/csteinmetz1/pyloudnorm}.} The final preprocessed and filtered GigaSpeech dataset contains a total of only 550 hours of speech.

For training the larger, 300M \model variant, we combine \textbf{1)} the aforementioned 550 hours of preprocessed GigaSpeech with \textbf{2)} the full LibriTTS dataset~\citep{libritts} (585 hours, 24 kHz) and a randomly subsampled 25\% of the English portion of Multilingual LibriSpeech (MLS) \citep{mls} ($\sim$10,000 hours, 24 kHz). All the data points were downsampled to 16 kHz. SpeechTokenizer has a temporal downsampling factor of 320, that is, for an input waveform sampled at 16 kHz it will produce quantized representations at 50 Hz.

For the single-speaker setup with \textit{synthetic data} where we start from the pretrained multi-speaker 300M \model model, we select \textit{one voice} from a well-known proprietary TTS provider\footnote{Not disclosed for confidentiality.} which was used in a real production system, and then set aside 270 generated utterances from real conversations (a total duration of 17 minutes) for evaluation, while the remaining synthetic utterances are used for training: there are two versions of training data spanning 7 hours and 10 hours of speech. The same preprocessing and filtering steps as with LibriTTS and MLS have been applied, including resampling to 16 kHz.



\sparagraph{Training Setup}
For both T2S and A2S components, we use $10,000$ warmup steps, learning rate is set to $5$e$-4$, and we rely on the AdamW optimizer~\citep{adamw} with $\beta_1 = 0.9$ and $\beta_2 = 0.98$. The learning rate is linearly decayed from $2 \times 10^{-4}$ to $0$.  For non-GigaSpeech utterances, we cut off the duration of each data point to 30 seconds. In both cases, training was carried out for $800$k steps with $8$ A10 GPUs in \texttt{bf16} precision.

\sparagraph{Inference Setup}
For the T2S component, we run multinomial sampling with temperature set to $0.7$ and $top\_k$ set to $210$. For A2S, we perform greedy sampling in the first level of acoustic tokens ($q_2$) and apply 16 steps of confidence-based sampling for all the remaining levels.\footnote{The hyperparameters were obtained via grid search over the following candidate values: temperature over the set $[0.3, 0.7, 1.0]$, $top\_k$ over $[60, 120, 150, 210]$ and inference steps for A2S over the set $[1, 4, 8, 16, 25]$. We performed inference on the validation set for all the possible hyperparameter combinations and selected the combination which minimized WER (see \S\ref{ss:metrics} on the details of how WER is computed).}


\subsection{Evaluation Setup and Metrics}
\label{ss:metrics}
We evaluate \model across four key dimensions of any multi-speaker (conversational) TTS system: \textbf{1)} speech intelligibility, \textbf{2)} voice preservation, \textbf{3)} reconstruction error measuring proximity to the reference ground truth speech, and \textbf{4)} prosody diversity and naturalness. Further, \textbf{5)} we assess inference efficiency reflecting the usability of the model in `real-time' production scenarios. The actual metrics per each evaluation dimension are described in what follows.

\sparagraph{Evaluation Data}
Since we care about conversational aspects of the model, we designed a particular evaluation set where each data point consists of two consecutive utterances pronounced by the same speaker: the first utterance always serves as the prompt both for T2S and A2S components of the model (see \S\ref{sec:methodology}). To create this evaluation dataset we measure the cosine similarity of the speaker embeddings extracted from consecutive utterances of the GigaSpeech development split. We retain only pairs of consecutive utterances where their cosine similarity is greater than 0.5. The final dataset comprises 227 \textit{(prompt, target)} pairs used for model evaluation.



\sparagraph{Speech Intelligibility: ASR Word Error Rate (WER)}
Following prior work~\citep{hayashi,mqtts}, in order to assess synthesised speech intelligibility we rely on Word Error Rate (WER) obtained by a state-of-the-art ASR system, Whisper\footnote{\url{https://github.com/openai/whisper}} \texttt{large-v2}. We convert text to lowercase and remove punctuation. 

\sparagraph{Voice Preservation: Speaker Similarity Score (SSS)}
Since \model is a multi-speaker model, we assess to what extent distinct speaker characteristics are preserved in synthesised speech, reporting the standard SSS scores. For measuring speaker similarity we employ WavLM \citep{wavlm}, a state-of-the-art self-supervised model according to the SUPERB benchmark~\citep{superb}.\footnote{\url{https://superbbenchmark.org}} We use released model parameters and supplementary scripts\footnote{\url{https://github.com/microsoft/UniSpeech}} for measuring speaker similarity. This consists of extracting features from the WavLM backbone model for both ground-truth and synthetic waveforms and measuring pairwise cosine similarity in the latent space.

\sparagraph{Reconstruction Error: Mel-cepstral Distortion (MCD)}
This metric, also established in prior work~\citep{hayashi,mqtts}, computes the reconstruction error between the ground truth speech and synthesis, with the fixed text and speaker. To time-align ground-truth and synthesized utterances we use dynamic time warping~\citep{dtw} and compute mean Euclidean distance between MEL-cepstral coefficients of the corresponding time-aligned audio frames. This measure characterizes distortion of acoustic information and is widely used to roughly evaluate speech quality, because it correlates well with human perception~\citep{mcd}.

\sparagraph{Prosody Diversity and Naturalness:  Fr\'{e}chet Inception Distance (FID)}
We again follow prior work~\citep{mqtts} for measuring prosody diversity and naturalness, reporting the FID scores~\citep{fid}. FID is an established objective metric in the evaluation of text-to-image generation models, originally created to measure realism and diversity of generated images, but its formalism makes it amenable also to a wider spectrum of generative models including the speech-oriented ones. More formally, it calculates the 2-Wasserstein distance between the collections of real/true speech segments and the corresponding synthesized/fake distributions, aiming to capture naturalness and diversity of the synthesised collection. This computation, of course, implies the use of a sufficiently large speech collection such as GigaSpeech used in our work. We use an open-source implementation of FID \footnote{\url{https://github.com/mseitzer/pytorch-fid}} score for our measurements following the MQTTS evaluation setup.

\sparagraph{Efficiency: Real-Time Factor (RTF)}
It is a standard metric of measuring the speed (or latency) of an audio processing (or any other processing) system. If we define $f(t_s)$ as the time required to process the input speech $s$ of duration $t_s$, RTF is computed as $f(t_s)/t_s$. Put simply, as a hardware-dependent metric, RTF tells how many seconds of speech are generated in one second of wall time. We used A10 and A100 NVIDIA GPUs for hardware tests. 

For the short sentence tests we used the following sentence: \textit{`Regrettably, we can't accommodate pets.'}, which is expected to last less than 3 seconds in a casual conversation. For the long sentence test, the following one is used: \textit{`Regrettably, we can't accommodate pets. However, we do permit assistance animals, provided they adhere to ADA regulations. For instance, if you're making arrangements for a stay at The Blue Finch Hotel in Naples, Italy, kindly ensure your service animal complies with this.'} This sentence should take more than 10 seconds to utter in a casual conversation.

\section{Results and Discussion}
\label{sec:results}
\label{ss:results}
\sparagraph{\model versus MQTTS}
We first aim to establish whether switching to fast parallel decoding hurts overall quality of conversational TTS synthesis. In this experiment, we compare the \textsc{small} \model variant (100M parameters) against the same-sized MQTTS model (as a state-of-the-art conversational TTS model) trained on the same GigaSpeech corpus (550 hours of preprocessed and filtered data, see \S\ref{ss:setups}). The main results are summarized in Table~\ref{tab:mqtts}. They reveal that, despite having much quicker, non-autoregressive inference, \model outperforms \mqtts on WER and MCD. As expected, while both models have reasonably high WER scores due to noisy and modestly sized training data (only 550 hours of training speech data in total), we note that \model is able to produce more diverse and more intelligible speech (offering reductions in WER of almost 2 points and higher FID scores). While slightly lower than those of MQTTS, speaker similarity scores still display high fidelity rates with production-ready potential.

\begin{table}[!t]
\def\arraystretch{0.999}
\centering
{\footnotesize
\begin{tabularx}{\textwidth}{l YYYY}
\toprule
\rowcolor{Gray}
\textbf{Model} & {\bf WER} $\downarrow$ & {\bf SSS} $\uparrow$ & {\bf MCD} $\downarrow$ & {\bf FID} $\downarrow$ \\
\cmidrule{2-5}
{\mqtts (100M)} & {14.2} & {\bf 0.682} & {9.568} & {19.690} \\
{\model-\textsc{small} (100M)} & {\bf 12.4} & {0.594} & {\bf 8.838} & {\bf 20.349} \\
\hdashline
{\model-\textsc{small} (100M), w/o SE} & {16.3} & {0.492} & {8.893} & {20.608} \\
\cmidrule{2-5}
{\model-\textsc{large} (300M)} & {11.9} & {0.549} & {8.671} & {19.675} \\
\bottomrule
\end{tabularx}
}%
\vspace{-0.5mm}
\caption{Comparison of \model and \mqtts with GigaSpeech training and test data (see \S\ref{ss:setups}). `w/o SE' is an ablation that does not use speaker embeddings for conditioning in the A2S component (\S\ref{ss:a2s}). We also show the results of the \model-\textsc{large} model for completeness (bottom row), but the reader should be aware that it cannot be directly compared to the other models as it was trained on a much larger (and different) training set (see \S\ref{ss:setups} again).}
\label{tab:mqtts}
\end{table}

\begin{table}[!t]
\def\arraystretch{0.999}
\centering
{\footnotesize
\begin{tabularx}{0.75\textwidth}{l YY}
\toprule
\rowcolor{Gray}
\textbf{Model} & {\em short} & {\em long} \\
\cmidrule{2-3}
{\mqtts (100M)} & {1.930} & {1.842} \\
{\model-\textsc{small} (100M)} & {\bf 0.133} & {\bf 0.133} \\
{\model-\textsc{large} (300M)} & {0.143} & {0.143} \\
\bottomrule
\end{tabularx}
}%
\vspace{-0.5mm}
\caption{Inference speed (RTFs, lower is better) on a single A100 GPU for two inference cases: \textit{short} and \textit{long} sentences (see \S\ref{sec:experiments}). Very similar RTFs with similar trends are obtained on an A10 GPU.}
\label{tab:speed}
\end{table}

Delving deeper into model failures, we observe that a significant portion of errors (in terms of WER) stems from the misspellings of proper nouns and homonyms. This suggests that using an ASR model with language model-assisted decoding can further reduce WER by favoring output sequences that contain out-of-distribution words such as proper nouns.

Table~\ref{tab:mqtts} also shows the scores of the 300M \model variant on the GigaSpeech test data, showcasing slight gains over the 100M variant in WER, MCD, and FID scores, but the reader should be aware that the 300M variant uses more abundant training data (beyond only preprocessed GigaSpeech) than \mqtts and \model-\textsc{small}.

\sparagraph{Inference Efficiency}
After establishing the relative synthesis quality, we investigate the efficiency aspect. Inference speed (RTF) for \mqtts and \model is provided in Table~\ref{tab:speed}. Coupled with the results from Table~\ref{tab:mqtts}, these results suggest huge benefits in terms of efficiency via non-autoregressive decoding, without compromising the generation quality: for instance, while a speech utterance of $\sim$10 seconds would require 18.4 seconds to process with MQTTS, which is infeasible in production, the 300M \model model requires only $\sim$1.4 seconds of processing time (a $14.5\times$ speed-up). Even a sentence with a duration of 3 seconds requires almost 6 seconds of processing with MQTTS, while the processing time is only 0.4 seconds with \model. The scores further reveal that the RTF scores with \model are not impacted at all by the expected output duration. Even more importantly, the reported RTFs indicate that the 3x larger model maintains almost the same and very competitive `production-friendly' inference speed while offering substantially higher quality synthesis. In sum, \model is able to maintain all the desirable qualities of MQTTS (i.e., conversational nature, parameter efficiency, sample efficiency) while simultaneously offering the ability to easily scale the model and increase its overall synthesis quality while maintaining unmatched inference speed. 

\begin{table}[!t]
\def\arraystretch{0.999}
\centering
{\footnotesize
\begin{tabularx}{\textwidth}{l YYYY}
\toprule
\rowcolor{Gray}
\textbf{Fine-tuning [hours]} & {\bf WER} $\downarrow$ & {\bf SSS} $\uparrow$ & {\bf MCD} $\downarrow$ & {\bf FID} $\downarrow$ \\
\cmidrule{2-5}
{0} & {3.9} & {0.533} & {24.799} & {6.314} \\
\hdashline
{7} & {4.7} & {0.578} & {31.426} & {5.653} \\
{10} & {4.5} & {0.576} & {33.251} & {5.677} \\
\bottomrule
\end{tabularx}
}%
\vspace{-0.5mm}
\caption{Single voice quality with different amounts of artificial data for single-speaker fine-tuning, starting from the \model-\textsc{large} (300M) multi-speaker checkpoint (first row of the table).}
\label{tab:empress}
\end{table}



\sparagraph{Single-Speaker Specialization}
Next, we take the \model-\textsc{large} (300M) multi-speaker model and evaluate it in a production scenario where the goal is to create a single `brand voice', a specialised TTS model that can output high-quality conversational TTS relying on a single speaker. We train the 300M variant on 7h and 10h of synthetic speech uttered in the same voice (see \S\ref{sec:experiments}). The results are provided in Table~\ref{tab:empress}. First, we note that even the multi-speaker 300M \model without any specialization exhibits very strong performance with low WER rates. Further fine-tuning on synthetic data yields slight losses in WER scores, but it offers gains in speaker similarity and FID scores, showing that the \model checkpoint can get fine-tuned even with synthetic data to better capture the single `brand voice'. If the goal is single-speaker specialization of a multi-speaker model checkpoint, we expect that further improvements might be achieved by (i) providing real human (instead of synthetic) training data uttered in the target voice, and (ii) scaling up training data: this warrants further investigation.

\begin{table}[!t]
\def\arraystretch{0.999}
\centering
{\footnotesize
\begin{tabularx}{\textwidth}{l YYYY}
\toprule
\rowcolor{Gray}
\textbf{Fine-tuning [hours]} & {\bf WER} $\downarrow$ & {\bf SSS} $\uparrow$ & {\bf MCD} $\downarrow$ & {\bf FID} $\downarrow$ \\
\cmidrule{2-5}
{0} & {4.7} & {0.445} & {28.761} & {6.436} \\
\hdashline
{7} & {4.1} & {0.502} & {34.703} & {5.813} \\
{10} & {4.5} & {0.504} & {33.020} & {5.829} \\
\bottomrule
\end{tabularx}
}%
\vspace{-0.5mm}
\caption{Ablation: A2S without speaker embeddings. Single voice quality with different amounts of artificial data for single-speaker fine-tuning, starting from the \model-\textsc{large} (300M) multi-speaker checkpoint (first row of the table); cf. Table~\ref{tab:empress} for the results with the full \model model.}
\label{tab:ablation}
\end{table}

\sparagraph{Ablation: A2S without Speaker Embeddings}
We also analyze the importance of including speaker embeddings in A2S training through an ablation. As revealed both in GigaSpeech experiments (Table~\ref{tab:mqtts}) and in single-speaker specialization experiments (compare the results in Tables~\ref{tab:empress} and \ref{tab:ablation}), the use of speaker embeddings has substantial positive impact on final performance, especially in terms of achieved fidelity (captured through the SSS scores). In the GigaSpeech evaluation, inclusion of speaker embeddings in fact positively affects all the evaluation metrics.

\sparagraph{On T2S and A2S}
We also analyzed \model through the lens of its two pivotal components: T2S and A2S. Our findings suggest that T2S is in fact the bottleneck of the entire system. While the A2S component typically performs robustly (modulo hyperparameter tuning) even in low-parameter setups (even lower than the 55M component used in \model 100M), further scaling down the T2S component yields learning instability and much deteriorated generation quality. Furthermore, with parallel decoding in A2S, the T2S component is now the main efficiency bottleneck: T2S processing consumes roughly 90\% of the total TTS processing time. This also means that relying on highly parameterized, large models (even with $\geq$ 1B parameters) for T2S as typically used in prior work \citep{speartts,soundstorm} is unsustainable for real-time production systems. This preliminary analysis suggests that future work should pay more attention to further enhancements of the T2S component, aiming to provide a better trade-off between performance and efficiency of T2S.

\section{Conclusion and Future Work}
\label{sec:conclusion}
We introduced \model TTS models, with the goal of creating and steering high-quality and production-friendly TTS systems that are conversational and highly parameter-, data-, and inference-efficient. To this end, we systematized and then adapted and extended recent advancements in Transformer-based TTS models, showing that it is possible to build a highly efficient TTS model that is up to almost 15x quicker at inference than previous state-of-the-art conversational TTS model~\citep{mqtts} while maintaining or even improving speech generation quality. We also showed that multi-speaker \model models can be effectively scaled up and specialised for particular voices via distillation with synthetic training data. 

We hope that the models shared with the community will serve as a solid departure and reference point for further developments of conversational and product-oriented TTS systems, aiming to focus on the interplay and trade-off between synthesis quality and efficiency. However, we have only scratched the surface of possibilities in this work. As indicated before, future work might look into other architectures and further improvements of the bottleneck text-to-semantics (T2S) component or could explore other parallel and efficient decoding strategies for TTS, e.g., speculative decoding~\citep{speculative}. Another avenue of future research might target the use of established parameter-efficient and modular designs~\citep{modulardl} for quicker and more efficient adaptation, or work on multilingual TTS~\citep{multilingualtts}, or on synthesis of very long utterances. Finally, our work indicated that there is still a large gap in high-quality conversational TTS data, so future research should also put more focus on such data collection and curation. 


\section{Acknowledgements}
We would like to thank our colleagues at PolyAI for many fruitful discussions and for their help with evaluation. We are also grateful to Michael Chen at PolyAI for the introduction and to Amazon Web Services for the computation donation which allowed us to run large-scale experiments.

\bibliography{references}
\bibliographystyle{iclr2023_conference}

\clearpage
\appendix
\section{Appendix}
\subsection{\model Model: Architecture}
\label{app:model_parameters}
The parameters of the T2S and A2S components, both coming in two different sizes, are summarized in Table~\ref{tab:parameters}. 

\begin{table}[!t]
\def\arraystretch{0.999}
\centering
{\footnotesize
\begin{tabularx}{\textwidth}{l cccYYY}
\toprule
\rowcolor{Gray}
\textbf{Variant} & {\# layers (enc)} & {\# layers (dec)} & {Hidden dim} & {FFN dim} & {Head dim} & {\# heads} \\
\cmidrule(lr){2-7}
{T2S-\textsc{small}} & {6} & {6} & {512} & {2,048} & {64} & {8} \\
{T2S-\textsc{large}} & {14} & {14} & {512} & {2,048} & {64} & {8} \\
\end{tabularx}
\begin{tabularx}{\textwidth}{l ccYYY}
\toprule
\rowcolor{Gray}
\textbf{Variant} & {\# Conformer layers} & {Kernel size} & {Hidden dim} & {FFN dim} & {\# heads} \\
\cmidrule(lr){2-6}
{A2S-\textsc{small}} & {3} & {5} & {1,024} & {1,024} & {8} \\
{A2S-\textsc{large}} & {8} & {5} & {1,024} & {1,024} & {8} \\
\bottomrule
\end{tabularx}
}%
\vspace{-0.5mm}
\caption{Architecture details for the two components of \model with different sizes.}
\label{tab:parameters}
\end{table}

\end{document}